\documentstyle[12pt]{article}
\begin{document}
\def\th{\theta}
\def\eps{\epsilon}
\def\vk{{\bf k}}
\def\de{\Delta}
\def\cj{{\Im}}
\def\kb{k_B\beta}
\def\beq{\begin{equation}}
\def\enq{\end{equation}}
\def\beqn{\begin{eqnarray}}
\def\eenq{\end{eqnarray}}
\def\pl{\parallel}
\baselineskip21pt
\begin{center}
{\large{\bf Effects of $c$-axis Hopping in the Interlayer 
            Tunneling \\Model of High-$T_c$ Layered Cuprates}}\\  
\vspace{0.8cm} 
{\bf Biplab Chattopadhyay$^{\star}$ and A. N. Das}\\
\vspace{0.2cm} 
{Saha Institute of Nuclear Physics, 1/AF Bidhannagar, 
      Calcutta - 700 064, INDIA}  
\end{center} 
\vspace{1cm}

\begin{abstract}
We consider the interlayer pair-tunneling model for layered 
cuprates, including an effective single particle hopping 
along the $c$-axis. A phenomenological suppression of the 
$c$-axis hopping matrix element, by the pseudogap in cuprate 
superconductors, is incorporated. At optimal doping, 
quantities characteristic to the superconducting state, such 
as the transition temperature and the superconducting gap   
are calculated. Results from our calculations are consistent 
with the experimental observations with the noteworthy point 
that, the superconducting gap as a function of temperature 
shows excellent match to the experimental data. Predictions 
within the model, regarding $T_c$ variation with interlayer 
coupling, are natural outcomes which could be tested further.  

\end{abstract}
 
\vspace{1ex} 
      
\noindent{PACS numbers: 74.80.Dm, 74.72.-h, 74.20.-z}  

\vfill

\noindent \rule{15.73cm}{0.1mm}\\
{\small $^\star$email: biplab@cmp.saha.ernet.in}; 

\newpage 
\noindent{{\bf 1. Introduction}}

A salient common characteristic feature of high temperature 
layered superconductors (HTLS) is the existence of well 
separated CuO$_2$ planes and the presence of strong electronic 
correlations in these planes. This characteristic, together 
with the observed interlayer contact, though very weak, 
often forms the basis for theoretical modelling of HTLS. 
Many of the to-date models of HTLS are intraplanar, 
focussing mainly on the electornic interaction in a single 
CuO$_2$ layer and ignoring, in the first approximation, the 
weak interlayer contact. But unlike these intraplanar 
models, the interlayer pair tunneling (ILPT) model due to 
Anderson \cite{pwand} is the one where 
interlayer contact has duly been incorporated and interlayer 
coupling plays an important role in this model to 
make it more compatible with the realistic HTLS materials 
than other existing models. 

The ILPT model of interacting electrons on coupled superconducting 
layers could naturally yield high transition temperatures ($T_c$) 
observed in the layered superconducting materials. In the ILPT 
model, single particle hopping along the $c$-direction was argued 
to be blocked owing to strong electronic correlations \cite{pwa22} 
and the tunneling of only pairs of carriers, caused by interlayer 
coupling, was considered. The idea behind pair tunneling is  
to amplify the pairing mechanism within a given layer, and as 
a consequence $T_c$ gets enhanced \cite{grpand}. 

It is important to note that, the interlayer hypothesis was  
based on certain experimental observations on underdoped 
cuprates, that is, the $c$-axis infrared conductivity is 
small, and the $c$-axis resistivity shows semiconducting 
behaviour in contrast to the $ab$-plane resistivity which 
is metallic \cite{infra}. These observations,  
which are signatures of the marginal presence of $c$-axis 
transport, led Anderson and co-workers to enforce complete 
suppression of interlayer single particle hopping (ISPH) in 
their ILPT model \cite{gr22and}. However, later development 
in sample preparation and the emergence of more accurate 
experimental techniques, made it possible to study the 
$c$-axis transport for wide range of dopings and it is observed 
that for sufficiently overdoped cuprates, both $c$-axis and 
$ab$-plane resistivities show similar temperature dependence 
\cite{expres} implying a three dimensional metallic behaviour 
of the material. In other words, for overdoped HTLS materials, 
$c$-axis transport is metallic-like. This view is further supported 
by the $c$-axis optical conductivity measurements on overdoped 
cuprates \cite{expres,uchida}, where the presence of Drude peak 
in the spectrum indicates the existence of interlayer single 
particle hopping. Also, band structure calculations yield 
substantial value for the interlayer single particle hopping 
matrix element for bilayer cuprates \cite{band}. 

Thus, guided by the to-date experimental results regarding the 
$c$-axis transport in variedly doped HTLS, we consider an extended 
or more generalized version of the ILPT model including an effective 
ISPH between the layers. Consideration of an {\em effective} ISPH 
is essential to remain content with the experimentally observed 
doping dependence of the $c$-axis transport that, the effective 
ISPH is strongest in the overdoped regime, decreases gradually 
through optimal doping and becomes negligible in the underdoped 
regime. In case of HTLS, the interrelation between the effects 
of electronic correlations and the variation of dopant concentration, 
also supports this doping dependence of effective ISPH. 
Effects of electronic correlations \cite{kuetal}, which inhibits 
ISPH within the Anderson's ILPT model, is strongest in highly 
underdoped cuprates leading to non-Fermi liquid characteristics 
\cite{scpwa}, but becomes weaker in the overdoped regime 
where a Fermi-liquid behavior is expected. Thus, it is suggestive
theoretically that the restriction on ISPH be relaxed in the
overdoped regime and ISPH be given due consideration. To remain 
consistent with these observations and arguments, we introduce a
probability factor $P$ to describe the effective ISPH, such 
that, $P$ becomes very small or negligible for highly underdoped
systems and increases with doping leading to significant values of 
the effective ISPH in the overdoped region.

An important recent development in HTLS is the observation of a
normal state pseudogap. This gap has been inferred from the NMR 
\cite{psnmr,william1}, optical conductivity \cite{homes,psoptc},
heat capacity \cite{loram} and transport data \cite{pstrans} in 
different cuprate materials. In addition, the angle resolved
photoemission spectroscopy (ARPES) experiments have shown that 
the symmetry of the normal state pseudogap is d-wave like 
\cite{psgap}, similar to that of the superconducting gap, and 
its magnitude ($E_g$) is large for underdoping, but falls off 
rapidly with the increase of doping \cite{william1,loram,psgap}. 
C. C. Homes {\it et al.}, by optical 
conductivity measurements on Y123 material \cite{homes}, found 
that for oxygen-reduced sample (underdoped), where $E_g$ is 
appreciable, the $c$-axis dc (zero frequency) conductivity
decreases with lowering of temperature, whereas at frequencies
well above the pseudogap the conductivity is temperature
independent. Similar evidences are also reported in experiments 
on other cuprates \cite{uchida,psoptc}. These findings 
reveal a clear correspondence between the
pseudogap magnitude and the suppression of $c$-axis conductivity.
Motivated by these facts we express the probability factor $P$
in terms of $E_g$ \cite{bipdas,dasil} in such a way, that the 
doping dependence of $E_g$ makes $P$ to follow the observed 
doping dependence of $c$-axis transport, i.e. the effective  
ISPH becomes small for underdoping and significant for overdoping.
Initially we choose an exponential form for the probability
factor $P=\,{\rm e}^{-E_g/T}$, which is very small
for underdoping (because of large $E_g$) ensuring a negligibly
small value of the effective ISPH. With the increase of doping,
$E_g$ falls off and $P$ increases, yielding significant values of
effective ISPH for the overdoped systems. Notice that, when
ISPH probability is finite, a fraction of the available charge
carriers will take part in ISPH, and only those particles, which are 
not participating in ISPH, will be available for pair tunneling. 
Since pair tunneling is a two particle process the probabilistic 
weight for such a process is taken as $(1-P)^2$. Here, in the 
generalised pair tunneling model, we assume that the interlayer 
tunneling of particles could occur via two channels: (i) single 
partile hopping and (ii) pair tunneling, and these two processes 
are complementary.
  
Having introduced the extended ILPT model, we do a mean-field 
analysis within the BCS approximation and obtain self consistent 
equations for the gap parameter and the chemical potential. 
Focussing mainly on the optimal doping situation (at which $T_c$ 
becomes maximum), we calculate 
$T_c$ and the superconducting gap $\Delta_\vk^{max}(T)$. Main 
results from our calculations are as follows. Phase diagram of 
the model ($T_c$ versus $\delta$) shows that the transition 
temperature $T_c$ at optimal doping increases with the increase 
of $E_g$. $T_c$ as a function of bare interlayer hopping 
($t_\perp$) shows upward or downward trend depending on the 
values of $E_g$. High values of the ratio of the 
superconducting gap to $T_c$ ($\sim 6-9$), as observed in HTLS 
\cite{exprat}, are recovered for realistic range of parameters. 
Gap variation $\Delta_\vk^{max}(T)/\Delta_\vk^{max}(0)$ as a 
function of scaled temperature $T/T_c$ shows qualitatively 
correct behaviour as in high-$T_c$ cuprates. A rigorous fit to 
the experimental gap-variation data \cite{gapexpt} is obtained 
with a $T$-linear choice of the probability factor $P$. 
 
The paper is organised as follows. In section-2 we present a 
detailed account regarding the formulation and justification of 
the extended ILPT model. Section-3 includes a brief presentation 
of the steps involved in the mean-field calculations leading 
to the gap equation. Results from our calculations are discussed 
in section-4 and section-5 contains a summary of results and 
some comments.  

\noindent{{\bf 2. Extended interlayer pair tunneling model: 
                  formulation and justification}}

The model Hamiltonian for the coupled bilayer complex 
\cite{dasil} is given by 
\begin{eqnarray}
H &=& \sum_{i,\vk,\sigma} (\epsilon_\vk - \mu)
    c_{\vk \sigma}^{(i)\dag} c_{\vk \sigma}^{(i)}  
  - \sum_{i,\vk,\vk^\prime} \left[ V_{\vk,\vk^\prime}\, 
    c_{\vk\uparrow}^{(i)\dag} c_{-\vk\downarrow}^{(i)\dag} 
    c_{-\vk^\prime\downarrow}^{(i)} c_{\vk^\prime\uparrow}^{(i)} 
    + h.c \right] \nonumber\\ 
 &-& \sum_{i\neq j,\vk} \left[ T_p^{\rm eff}(\vk)\, 
       c_{\vk\uparrow}^{(i)\dag} c_{-\vk\downarrow}^{(i)\dag} 
       c_{-\vk\downarrow}^{(j)} c_{\vk \uparrow}^{(j)} + h.c \right] 
     + \sum_{i\neq j,\vk,\sigma} \left[ t_\perp^{\rm eff}(\vk)\, 
       c_{\vk \sigma}^{(i)\dag} c_{\vk \sigma}^{(j)} + h.c \right]  
\end{eqnarray}
which is similar to that of pair tunneling model \cite{pwand},  
except the last term that accounts for the interlayer single 
particle hopping. Here, $c_{\vk\sigma}^{(i)\dag}$ 
($c_{\vk\sigma}^{(i)}$) is the fermion creation (annihilation) 
operator with momentum $\vk$ and spin sigma, $i\,(=1,2)$ is the layer 
index, $V_{\vk,\vk^\prime}$ is the pairing potential forming 
Cooper pairs in the $ab$-plane and $\mu$ is the chemical potential. 
The $ab$-plane band dispersion $\epsilon_k$ is taken to be that 
of Bi2212, obtained from a six parameter tight binding fit 
$[t_0, t_1, t_2, t_3, t_4, t_5]$ 
= $[0.131,-0.149,0.041,-0.013,-0.014,0.013]\,eV$ to the ARPES data  
\cite{norman}. This six parameter band dispersion, used elsewhere 
\cite{biplab}, shows flat bands in the Brillouin zone and the 
corresponding density of states (DOS) has a power law singularity 
known as extended van Hove singularity. This is characteristic 
to the high-$T_c$ cuprates \cite{evhs}. $T_p^{\rm eff}(\vk)$ and 
$t_\perp^{\rm eff}(\vk)$ represent effective matrix elements for 
pair tunneling and single particle hopping respectively between 
the layers which involve the probability factor $P$. Effective 
ISPH is taken to be $t_\perp^{\rm eff}(\vk) = t_\perp^b(\vk) P$, 
where $t_\perp^b(\vk)=t_\perp((\cos k_xa-\cos k_ya)/2)^2$ is the 
$\vk$-dependent ISPH as predicted by the band structure calculations 
\cite{band} with $t_\perp$ being the bare ISPH matrix element 
($a$ is the lattice constant). The bare pair tunneling, following 
the original ILPT model \cite{grpand}, is taken as   
$T_p(\vk) = [(t_\perp^b(\vk))^2/|t_1|]$ where $t_1$ is the 
nearest neighbour hopping matrix element of the $ab$-plane 
band dispersion \cite{norman}, and the effective pair tunneling,  
as mentioned earlier, is represented by 
$T_p^{\rm eff}(\vk) = T_p(\vk)\,(1-P)^2$.  

Regarding the origin of the pseudogap in cuprates no concensus 
has been reached so far. Some authors believe that it is precursor
to the supreconducting gap \cite{precur} and it represents the 
pairing energy of preformed pairs without phase coherence. 
Absence of phase coherence among the preformed pairs, is 
believed to be due to the strong phase fluctuations in the 
quasi two dimensional systems, and superconductivity sets in 
at the phase locking transition temperature of these pairs. 
However, such a scenario is contradicted by recent experiments 
\cite{william2}, which observe absence of any isotope effect
in $E_g$ even though there remains an isotope effect in $T_c$, 
suggesting that the interactions responsible for superconductivity 
and the pseudogap are independent and the pseudo gap cannot be 
attributed to short-range superconducting pairing correlations. 
Similar views are reflected by M. Suzuki {et al.} by tunneling 
spectroscopy measurements on Bi2212 material \cite{suzgrp}. These 
experiments also found that the normal state pseudogap coexists 
with the superconducting gap below $T_c$. 

A second school of thought pertains to the strong correlation 
viewpoint for HTLS \cite{pwa33}, where spin and charge degrees 
of freedom of electrons are decomposed into spinons and holons.  
For underdoped systems, below a temperature $T_{\rm RVB}$, the 
spinons are paired as Resonating Valence bond (RVB) \cite{andzou} 
in the singlet state, and $T_{\rm RVB}$ is much higher than the 
superconducting transition temperature \cite{fukuyama}. The 
pseudogap may be interpreted as the spin excitation
gap over the RVB singlet state. Within this picture, a holon has to
combine with a spinon to form a real hole which can hop between the
layers. Cosequently, for underdoped sytems where spinons are paired
below $T_{\rm RVB}$, the ISPH is heavily suppressed. Such a picture, 
which forms the  basis of Anderson's pair tunneling model,
also motivates us to assume that charge carriers, which 
are available for ISPH, decreases with $E_g$. This phenomenon
(for the case of short-range RVB)
could be represented by an exponential probability factor 
$P=\exp(-E_g/T)$. Exponential form of $P$ signifies that the particles
have to be activated above the gap $E_g$ to be available for ISPH. 
This activation is favoured as $T$ goes up or $E_g$ gets 
lowered, whereas an increase in $E_g$ would degrade the 
activation and lead to the suppression of ISPH. Notice 
that the form of $P$, as it stands, could mimic the 
observed doping dependence of $c$-axis transport by virtue 
of the parameter $E_g$, that is, $c$-axis transport is 
small for underdoping ($E_g$ large), gradually increases 
through optimal doping ($E_g$ gets lowered) and becomes 
significant for overdoping ($E_g$ very small).   

The phenomenological behaviour of $P$, and consequently that of 
the effective ISPH, is also supported by certain existing ideas 
and results in HTLS. Kumar and Jayannavar \cite{kujay}, while 
trying to explain the power law dependence of the 
semiconducting-like $c$-axis resistivity ($\rho_c$) of HTLS, 
showed that the low temperature up-turn of $\rho_c$ could 
come from the $T$-dependence of the interlayer coupling 
$t_\perp$ which gets renormalized as a power of temperature 
$t_\perp^{\rm eff} = t_\perp T^\alpha$ 
(Eq.(7) of Ref.\cite{kujay}) where $\alpha$ is an exponent 
of the order of unity. This renormalization of $t_\perp$, 
which is an adiabatic modification, was originally derived by 
A. J. Leggett {\it et al.} considering a coupling of the slow 
interplanar electron tunneling to some bosonic degrees of 
freedom \cite{legget}. It has also been established that the 
strong electronic correlations, present in the CuO$_2$ planes 
of HTLS in the form of strong intraplane electron-electron 
scattering, could lead to the complete blocking of single 
particle tunneling between CuO$_2$ layers at zero temperature 
($T=0$), but the tunneling of electron pairs remain 
uninhibited \cite{kuetal}. These views, if translated in the 
language of our model, would mean that at $T=0$, $P=0$ and 
$t_\perp^{\rm eff}=0$ but $T_p^{\rm eff}\neq 0$, whereas at 
any finite temperature ($T\neq 0$) $P\neq 0$ and 
$t_\perp^{\rm eff}$ is nonzero. Precisely, the effective ISPH 
and the effective pair tunneling matrix elements within our 
model show similar temperature dependence. 

Thus, our proposed model for HTLS is consistent with the 
experimental observations and theoretical ideas/results 
regarding the $c$-axis electrical transport. Effects of 
intraplanar correlations, which is strongest in the 
too-underdoped region ($E_g\to \infty$), can be thought of 
as coming to picture in our model as blocking the ISPH 
at any temperature \cite{kuetal} in the limit of very small 
doping. As the doping is increased, intraplanar correlations 
become weaker ($E_g$ decreases) and ISPH becomes more and more 
favourable at nonzero temperatures, however at $T=0$, ISPH 
remains completely blocked at any doping. Thus, in the limit 
$E_g\to \infty$ or at $T=0$, the probability factor $P$ 
vanishes leading to $t_\perp^{\rm eff} = 0$ and our model 
becomes equivalent to the original ILPT model, and in this 
sense the proposed model is more general. 

It is important to note, that the presence of ISPH gives rise to  
bilayer splitting, which however has not been observed in Bi2212 
system \cite{bilayer}, particularly at low temperatures. In this 
light, we need to check the validity of our proposed model. A 
study of the electronic density of states (DOS) of our model 
\cite{bipdas} reveals that for optimally doped or underdoped systems, 
when $E_g\geq T_c$, bilayer splitting is absent at low temperatures
$T\leq 20\,K$ and within an energy resolution of $1\, meV$.  
This is because of strong suppression of ISPH by $E_g$. At higher 
temperatures, broadening of the quasiparticle states would be 
important to obscure such splitting in real systems. A detailed 
calculation of the ARPES intensity curves within our proposed 
model clearly establishes that bilayer splitting would not be 
observable at low temperatures $\sim 13\,K$ as in actual 
experiment of Ref.\cite{bilayer}. Even at a higher temperature 
$\sim 40\,K$ and with a resolution much better than that in 
experiment, splitting remains absent \cite{bipdas}. These 
outcomes further embolden the consistency of our proposed 
model with experimental results on HTLS.  

\noindent{{\bf 3. Superconducting gap equation within the 
                  BCS approximation}} 

Mean field decoupling of the four fermion terms in the 
Hamiltonian of Eq.(1) yields 
\begin{eqnarray}
H &=& \sum_{i,\vk,\sigma} (\epsilon_k - \mu) 
      c_{\vk \sigma}^{(i)\dag} c_{k \sigma}^{(i)} - 
      \sum_{i,\vk} \left[\Delta_{\vk}\, c_{\vk \uparrow}^{(i)\dag} 
      c_{-\vk \downarrow}^{(i)\dag} + h.c \right] \nonumber\\
  &+& \sum_{i\neq j,\vk,\sigma} \left[t_\perp^{\rm eff}(\vk)\, 
      c_{\vk \sigma}^{(i)\dag} c_{\vk \sigma}^{(j)} + h.c \right] 
\end{eqnarray}
with the gap parameter being defined as  
\begin{equation}
{\Delta}_\vk = {\Delta}_{i,\vk} = \sum_{\vk^\prime} 
  V_{\vk,\vk^\prime}\, \langle c_{-\vk^\prime \downarrow}^{(i)} 
  c_{\vk^\prime \uparrow}^{(i)}\rangle  
  + T_p^{\rm eff}(\vk)\, \langle c_{-\vk \downarrow}^{(j)} 
  c_{\vk \uparrow}^{(j)}\rangle 
\end{equation}
Layer index $(i,j)$ are equivalent, since by symmetry, in-plane
pairing is identical in both the layers. Presence of interlayer 
single particle hopping produces two quasiparticle bands 
$E_\vk^- =\sqrt{\{\epsilon_\vk - \mu-t_\perp^{\rm eff}(\vk)\}^2 
+ \Delta_\vk^2}$\, and 
\,$E_\vk^+ =\sqrt{\{\epsilon_\vk - \mu+t_\perp^{\rm eff}(\vk)\}^2 
+ \Delta_\vk^2}$.  

Self consistent equations, for chemical potential and the 
superconducting gap are obtained as 
\begin{equation}
1-\delta = 1-\frac{1}{N} \sum_\vk \left(\epsilon_\vk-\mu -t_\perp^{\rm eff}\right) 
	   \chi(E_\vk^-)
	 - \frac{1}{N} \sum_\vk \left(\epsilon_\vk-\mu +t_\perp^{\rm eff}\right) 
	   \chi(E_\vk^+)
\end{equation}
and 
\begin{equation}
\Delta_\vk = \frac{\displaystyle\sum_{\vk^\prime} \Delta_{\vk^\prime}\, 
		V_{\vk,\vk^\prime} 
		\left(\chi(E_{\vk^\prime}^-) + 
                      \chi(E_{\vk^\prime}^+)\right)\!\!\bigg/2}
	       {1 - T_p^{\rm eff}(\vk) 
		\left(\chi(E_\vk^-)+\chi(E_\vk^+)\right)\!\!\bigg/2} 
\end{equation}
where $\chi(E_\vk^\pm)=\frac{1}{2E_\vk^\pm}
\tanh\left(\frac{\beta E_\vk^\pm}{2}\right)$, $\beta = 1/T$ (in 
a scale of k$_B$=1),   
$\delta=1-n$ with n being the number of electrons per site and 
$N$ is the total number of lattice sites. The pairing interaction    
$V_{\vk,\vk^\prime}$ is separable as 
$V_{\vk,\vk^\prime} = V\,\eta_\vk\,\eta_{\vk^\prime}$, which also 
makes the $\vk$-dependence of $\Delta_\vk$ to be shoved in the 
symmetry factor $\eta_\vk$. Finally, the expression for the 
superconducting gap (from Eq.(5)) becomes 
\begin{equation}
{1\over {4V}} = \frac{1}{N} \sum_{\vk} 
	     \frac{\eta_\vk^2\,\left(\chi(E_\vk^-) + 
                   \chi(E_\vk^+)\right)\!\!\bigg/2} 
	     {1 - T_p^{\rm eff}(\vk)\left(\chi(E_\vk^-) + 
                   \chi(E_\vk^+)\right)\!\!\bigg/2} 
\end{equation}
									 
In our calculations, the $d_{x^2-y^2}$ symmetry of the pairing 
state is considered because of growing evidence of the same in 
high-$T_c$ cuprates \cite{dwpap1,dwpap2} and this implies the 
symmetry factor to be $\eta_\vk = (\cos k_xa -\cos k_ya)/2$. 
The parameter $V$, in our case, is the nearest neighbour 
attractive interaction. In experiments, momentum dependence 
of the normal state pseudogap is found to be of $d_{x^2-y^2}$ 
symmetry \cite{psgap}, and we take 
$E_g(\vk)=E_g|\cos k_xa -\cos k_ya|$. This introduces a $\vk$ 
dependence in the probability factor 
$P(\vk)=\exp(-E_g(\vk)/T)$.   

\noindent{{\bf 4. Results and discussions}}  
 
We numerically solve the self consistent equations (4) and (6) 
for the superconducting gap parameter and the chemical potential. 
Basically, we focus on the properties characterizing the 
superconducting state, since in this state the quasiparticle 
picture is valid at any doping. We calculate the superconducting 
gap $\Delta_\vk^{max}(T)$ and the transition temperature $T_c$ 
(at which the gap magnitude drops to zero) and also study their 
variations with model parameters. 
   
\noindent{\em 4.1. Transition temperature:} 

In Fig.1, we plot transition temperature $T_c$ as a function 
of $\delta$, for various values of the normal state pseudogap 
$E_g$ as shown explicitly. For $E_g=0$, where ISPH takes its full 
value ($P=1$) and the pair tunneling is absent, $T_c$ has two peaks. 
This is a consequence of the splitting of DOS due to the presence 
of unsuppressed ISPH between the layers \cite{bipdas,dasil}. 
As $E_g$ increases, the ISPH gets suppressed and the two 
peak form gradually merge into one peak, yielding the maximum 
transition temperature ($T_c^m$) at optimal doing. For realistic 
values of $E_g$, the one peak form persists which is the case 
in experiments. We choose the bare value of $t_\perp=40\,meV$,  
in accordance with the band structure calculations \cite{band} 
suggesting $t_\perp/|t_1|\sim 0.2-0.3$. Interaction parameter 
is fixed at $V=70\,meV$ which makes $T_c^m \sim 100\,K$ for 
$E_g\sim T_c^m$. This value of $V$ is kept fixed throughout in this 
communication. Clearly, for fixed values of $V$ and $t_\perp$, 
$T_c^m$ increases with the increase of $E_g$ and 
highest value of $T_c^m$ is obtained in the Anderson limit 
($E_g=\infty$) where ISPH is completely suppressed giving 
vent to only pair tunneling between the layers. It should 
be noted that $E_g$ is kept fixed for every curve in Fig.1, 
whereas, in reality it should change with doping. Here, we 
are interested in quantities at optimal doping (as in Fig.2 
and Fig.3) and hence consider different fixed $E_g$ values 
which might represent different cuprate materials. Increase 
of $T_c^m$ with $E_g$ is consistent with the scaling behaviour 
of the pseudogap \cite{william1}, according to which, at a 
fixed doping the ratio $E_g/T_c^m$ should ideally fall on 
the same point for different cuprate materials. 

Variation of $T_c^m$ as a function of $t_\perp$ is shown 
in Fig.2. Numerical values of $E_g$ for different plots are 
(A,\,B,\,C,\,D,\,E) = (2,\,4,\,6,\,8,\,$\infty$) in $meV$. 
For small $E_g$, $T_c^m$ decreases with increasing $t_\perp$. 
Increase of $E_g$ makes interlayer pair tunneling more 
probable. Consequently, $T_c^m$ rises with $t_\perp$ and in 
the Anderson limit $T_c^m$\, grows rapidly with $t_\perp$. 
It may be noted that regarding the role of the interlayer 
coupling on $T_c$, there is mixed experimental evidence 
(see Ref.\cite{expres}, page 141). Our results show that 
$T_c^m$ may increase or decrease or may even remain unaffected 
with the increase of interlayer coupling $t_\perp$, depending 
upon the values of $E_g$. This is a prediction of our 
calculations and a systematic study of the variation 
of $T_c$ with $t_\perp$ for differently doped or different 
class of cuprate materials (with changing $E_g$) is 
called for. A possible mechanism to vary $t_\perp$ could 
be the application of pressure along $c$-axis. 
\vskip25pt

\noindent{\em 4.2. Superconducting gap parameter:} 

Two important issues relating the superconducting gap, which 
figures frequently in the HTLS literature, are the ratio of 
the superconducting gap to $T_c$ and the temperature variation 
of the gap. Superconducting gap-ratio (gap-width) 
$\Delta_\vk^{max}(0)/T_c$, as a function of $t_\perp$ for 
different $E_g$, is presented in Fig.3. Alphabetic labels 
correspond to different $E_g$ as written in the figure 
(similar to the values of $E_g$ as in Fig.2). In the Anderson 
limit, gap-ratio could reach the value $\sim 4.5$ for $t_\perp$ 
within realistic range ($\sim 30-45\,meV$). But, for finite 
$E_g$, the gap-ratio within this range of $t_\perp$ increases 
as $E_g$ is lowered. For example, with $E_g=4\,meV$ one could 
get the gap-ratio as high a value as $\sim 6$ for 
$t_\perp\sim 45\,meV$. To understand this, note that for any 
finite $E_g$ the probability factor $P\rightarrow 0$ as 
$T\rightarrow 0$. This situation corresponds to the Anderson 
limit where ISPH is completely suppressed. Hence at $T=0$, 
$\Delta_\vk^{max}(0)$ is quantitatively same for any $E_g$ 
whatsoever for a fixed $t_\perp$. But for $T\neq0$, as $E_g$ 
is lowered, the suppression of ISPH gets weaker and 
correspondingly $T_c$ falls off from its value in the Anderson 
limit which is evident from the plot in Fig.2. Consequently, 
the ratio $2\Delta_\vk^{max}(0)/T_c$ is small for $E_g=\infty$ 
and increases as $E_g$ is decreased. 
  
Finally, We study the temperature variation of the superconducting 
gap. In Fig.4 we plot $\Delta_\vk^{max}(T)/\Delta_\vk^{max}(0)$ 
as a function of reduced temperature $T/T_c$ (dashed lines), for 
different $E_g$ values shown. Here $t_\perp=40\,meV$ and the 
doping is at the optimum level. The solid line represents 
results for BCS superconductors and solid square symbols are 
experimental data \cite{gapexpt} for Bi-cuprate. Clearly, the 
locus of experimental data is well below the BCS-curve. Results 
from our calculations are all below the BCS-curve, but  high values 
of $E_g$ makes them closer to the BCS-curve, with the closest one 
being that in the Anderson limit. For small $E_g$ our calculations  
are close to the experimental data. However, our results with   
$P=\exp(-E_g/T)$ could not match the experimental data for the 
whole range of $T/T_c$. Maximum mismatch is observed in 
the low temperature region where experimental data registers 
a near linear fall from its $T=0$ saturation. 

Anticipating that the linear fall of $\Delta_\vk^{max}(T)$ 
in the vicinity of $T=0$ could come from a linear increment 
of effective ISPH, we consider another possible choice of $P$ 
as  
\begin{equation}
P = {{T}\over{E_g + T {\rm e}^{-E_g/T}}} 
\end{equation} 
Note that the choice of $P$ is not unique. While making 
a choice, the conditions to be satisfied are, $P\propto T$ 
for $E_g\gg T$ and $P=1$ for $E_g=0$, which are ensured in Eq.(7). 
Results, with linear-$T$ dependent $P$ and $t_\perp=40\,meV$, 
$E_g=8\,meV$ \cite{noteeg}, are presented in the inset of 
Fig.4 as a dashed line. The matching with experimental data 
(solid squares) is excellent. To our knowledge, this is the 
first ever calculations that correctly reproduces experimental 
data, affirming the merits of the model under consideration. 
Use of $P$ from Eq.(7) does not qualitatively change other 
results discussed earlier, rather yields higher values of 
the gap-ratio than those obtained with exponential $P$ for 
the same set of parameters. As for example, $t_\perp=40\,meV$ 
and $E_g=4\,meV$ yields a gap-ratio 7.1 for the present case. 

The $T$-linear form of the probability factor could 
have phenomenological support from the work by Leggett  
{\it et al.}, where the renormalization of $t_\perp$ involves 
a power law dependence of temperature as $T^\alpha$ with 
$\alpha$ being of the order of unity. Arguments could also be 
given within the spin-charge separation picture of Anderson 
and coworkers, where spinons and holons are the quasiparticles 
in a layer, and a holon has to combine with a spinon to form 
a real hole that can hop to another layer. Thus, within this 
picture, $c$-axis transport is proportional to the spinon 
density, and since the density of spin excitations grows as $T$ 
within the (long-range) RVB model \cite{andzou}, the $c$-axis 
transport is expected to increase linearly with $T$. These 
views, at least on the phenomenological level, justifies 
the form of $P$ as in Eq.(7).  

So far, we have presented results within the extended ILPT model
incorporating the pseudogap as a phenomenological parameter which
affects only the out-of-plane charge transport (suppresses the ISPH), 
and our focus remained on the optimally doped situation (fixed $E_g$).
But in actual case, the pseudogap should also affect the in-plane
charge dynamics and should vary with doping as observed in 
experiments \cite{william1,loram,psgap}. Regarding the
in-plane effects of $E_g$, based on the
analysis of experimental pseudogap data for
several cuprates, it has been suggested that, the pseudogap 
suppresses the in-plane spectral weight or single particle DOS
\cite{william1}, and to this effect, the total measurable spectroscopic
gap on the Fermi surface (FS) would be an additive combination
of the superconducting gap and the pseudogap. Following this idea,
we modify the quasiparticle band enegies as $E_\vk^\pm =
\sqrt{\{\epsilon_\vk -\mu \pm t_\perp^{\rm eff}(\vk)\}^2
+ \Delta_\vk^2 +E_g(\vk)^2}$ and do preliminary calculations  
incorporating the variation of $E_g$ with doping as in
Ref\cite{william1}. Our results show that, $T_c$ in this case, 
falls off rapidly towards underdoping because of the rapid 
suppression of spectral weight by high values of $E_g$. We also 
find that, the total spectroscopic gap at zero temperature ($T=0$) 
shows non-trivial doping dependence (different from $T_c$) as
noted by tunneling experiments on HTLS \cite{renner}. Around 
the optimal doping, in-plane effects of $E_g$ does not change 
the qualitative features of the superconducting gap or the 
transition temperature \cite{unpub}, and the conclusions drawn 
in this paper remains unaffected. However, to have better
understanding about the effects of $E_g$ on the $ab$-plane 
charge dynamics, and to draw definite conclusions regarding 
the behaviour of $T_c$ and gap parameter as a function of 
doping within the DOS-suppression picture, detailed 
calculations would be necessary. Results on these matters 
will be presented in future communications.  

\noindent{{\bf 5. Summary and comments}} 

In this communication, we consider an extension of the interlayer 
pair tunneling model including an effective interlayer single 
particle hopping and calculate the superconducting gap as well 
as the mean-field $T_c$. Here, we briefly state the results from 
our calculations. Increase of $T_c^m$ with the increase of $E_g$ 
is in qualitative agreement with the scaling behaviour of the 
pseudogap. Within the model, high values for the gap-ratio are 
obtained and the experimental gap-variation data are reproduced 
with a $T$-linear probability factor for the effective ISPH.  
An outcome of our calculations, that $T_c$ may increase or 
decrease as a function of interlayer coupling depending on $E_g$, 
comes as a natural prediction which could be put to test. Thus, 
several contentious issues of high-$T_c$ cuprates are explained 
within the extended interlayer pair tunneling model and new 
predictions are made.    

Certain comments regarding the phase diagram and the extended 
ILPT model in general, would be relevant at this point. 
In Fig.1 we find that for the no doping ($\delta=0$) situation  
a finite $T_c$ is obtained, which is not the case in real 
layered cuprate materials. This is due to the non-consideration 
of the effect of the strong on-site repulsions (strong correlations) 
on the kinetic energy term removing any double occupancy. The   
inconsistency can be removed by calculating $T_c$ within a $t-J-V$ 
model where hopping is restricted in a space with no double occupancy   
\cite{andas}. But the point is that, in both of these analyses 
(in the limit of weak and strong correlations), properties 
characterising the superconducting state would remain unaltered. 
Moreover, we would like to note that, in our extended ILPT model,  
strong intraplanar correlations do exist, which plays role only to 
suppress the single particle hopping between CuO$_2$ layers. This is 
similar in spirit to the views adapted by Anderson and collaborators 
relating to the original ILPT model \cite{grpand}.     

In the context of ILPT model, Chakravarty and Anderson showed that, 
in the limit of very small temperature or ($T\to 0$), the phenomenon 
of the blocking of ISPH is exhibited by those HTLS materials 
characterized by non-Fermi liquid behaviour \cite{scpwa}. We would 
like to note, that non-Fermi liquid behaviour, as mentioned in the
introduction, is the characteristic
of too-underdoped HTLS ($E_g$ very large or $E_g\to \infty$) and the 
original ILPT model, which is a special case of our extended model, 
is valid for very little doped HTLS. For overdoping, on the other 
hand, HTLS show Fermi liquid like characteristics and the $c$-axis 
charge transport becomes more like that in the $ab$-plane. This 
implies that, for overdoped HTLS, ISPH gains significance and
an extension of ILPT model, as considered in this communication, 
becomes essential. Thus one finds that the extended ILPT model 
is applicable to HTLS for a wide range of doping starting from 
too little to very high, whereas the original ILPT model is 
applicable only in the underdoped region. Furthermore, comparison 
of our results with those from experiments on HTLS show that, 
the extended ILPT model captures HTLS phenomenology better than 
the original ILPT model. 

\baselineskip20pt
\noindent {\bf Acknowledgements:} 
 
It is a pleasure to thank B. Ghosh, S. Shenoy and C. Kawabata for 
useful discussions. One of us (B.C) thanks CSIR, Government 
of India, for a Senior Research Associateship (Pool Scheme).

\newpage
\baselineskip22pt

\noindent{\Large{\bf Figure captions:}}  

\begin{itemize}

\item[Fig.1.] Mean-field superconducting transition temperature ($T_c$) 
  as a function of dopant concentration ($\delta$) for different 
  values of $E_g$ (magnitude of the normal state pseudogap) as 
  shown in the figure. Other parameters are chosen to be 
  $V=70\,meV$ and $t_\perp=40\,meV$. Anderson limit corresponds 
  to $E_g=\infty$. 
\vskip 0.3 cm

\item[Fig.2.] Plot of $T_c^m$ as a function of bare interlayer single 
  particle hopping $t_\perp$, for different $E_g$ values denoted 
  by alphabetic labels (A,\,B,\,C,\,D,\,E) = (2,\,4,\,6,\,8,\,$\infty$) 
  in $meV$. Varied behaviour of $T_c^m$ flow with $t_\perp$ is 
  apparent from the curves.  
\vskip 0.3 cm

\item[Fig.3.] Maximum value of the gap-ratio ($2\Delta_\vk^{max}(0)/T_c$) 
  is plotted as a function of $t_\perp$ at optimal doping. Alphabatic 
  labels corresponding to different $E_g$ values are the same as in Fig.2 
  and are listed in the inset.   
\vskip 0.3 cm

\item[Fig.4.] Finite temperature gap scaled to its zero temperature value  
  ($\Delta_\vk^{max}(T)/\Delta_\vk^{max}(0)$), as a function of 
  reduced temperature ($T/T_c$). Solid line is the BCS-form, solid 
  square symbols are experimental data from Ref.\cite{gapexpt} and 
  dashed lines are from our calculations for different $E_g$ shown. 
  [Inset: Dashed line is from our calculations with the renormalizing  
  factor as in Eq.(7). Solid squares are experimental data and solid 
  line is the BCS-form as in the main figure].  

\end{itemize} 


\newpage

\begin{thebibliography}{999} 

\bibitem{pwand}{P. W. Anderson {\it et al.}, in {\it Superconductivity}, 
  Proceedings of the ICTP spring college in 1992, eds. P. Butcher and 
  Y. Lu (World Scientific, Singapore).}  

\bibitem{pwa22}{P. W. Anderson , Science {\bf 256} (1992) 1526.}  

\bibitem{grpand}{S. Chakravarty, A. Sudbo, P. W. Anderson and 
  S. Strong, Science {\bf 261}, 337 (1993).}

\bibitem{infra}{K. Tamasaku, Y. Nakamura and S. Uchida, 
  Phys. Rev. Lett. {\bf 69}, 1455 (1992); B. Batlogg, in 
  {\it High Temperature Superconductivity}, eds. K. S. Bedell 
  {\it et al.} (Addison-Wesley, 1990).}

\bibitem{gr22and}{A. Sudbo {\it et al.}, Phys. Rev. B {\bf 49}, 
  12245 (1993); L. Yin {\it et al.}, Phys. Rev. Lett. {\bf 78}, 
  3559 (1997); P. W. Anderson, Science {\bf 279}, 1196 (1998).}   

\bibitem{expres}{S. L. Cooper and K. E. Gray, in {\it Physical 
  Properties of High Temperature Superconductors IV}, 
  ed. D. M. Ginsberg (World Scientific, Singapore, 1994).}

\bibitem{uchida}{S. Uchida, K. Tamasaku and S. Tajima, Phys. Rev. B 
  {\bf 53}, 14558 (1996) and refrences therein.}   

\bibitem{band}{O. K. Andersen, A. I. Liechtenstein, O. Jepsen and 
  F. Paulsen, J. Phys. Chem. Solids {\bf 56}, 1573 (1995).}

\bibitem{kuetal}{N. Kumar, T. P. Pareek and A. M. Jayannavar, 
  Phys. Rev. B {\bf 57}, 13399 (1998).}

\bibitem{scpwa}{S. Chakravarty and P. W. Anderson, Phys. Rev. Lett. 
  {\bf 72}, 3859 (1994).}   

\bibitem{psnmr}{M. Takigawa {\it et al.}, Phys. Rev. B {\bf 43}, 
  247 (1991).}  

\bibitem{william1}{G. V. M. Williams, J. L. Tallon, R. Michalak and 
  R. Dupree, Phys. Rev. B {\bf 54}, 6909 (1996); G. V. M. Williams 
  {\it et al.}, Phys. Rev. Lett. {\bf 78}, 721 (1997).}  

\bibitem{homes}{C. C. Homes {\it et al.}, Phys. Rev. Lett. {\bf 71}, 
  1645 (1993).}  

\bibitem{psoptc}{A. V. Puchkov {\it et al.}, Phys. Rev. Lett. {\bf 77}, 
  3212 (1996).}  

\bibitem{loram}{J. W. Loram {\it et al.}, J. Supercond. {\bf 7}, 
  243 (1994).}  

\bibitem{pstrans}{ B. Batlogg {\it et al.}, Physica C {\bf 235-240}, 
  130 (1994).}  

\bibitem{psgap}{H. Ding {\it et al.}, Nature {\bf 382}, 512 (1996); 
  A. G. Loeser {\it et al.}, Science {\bf 273}, 325 (1996).}  

\bibitem{bipdas}{B. Chattopadhyay and A. N. Das, Phys. Lett. A 
  {\bf 246}, 201 (1998); A. N. Das and B. Chattopadhyay, Physica C 
  {\bf 308}, 226 (1998).}  

\bibitem{dasil}{A. N. Das and S. Sil, Physica C {\bf 299}, 83 (1998).}  

\bibitem{exprat}{T. Hasegawa, H. Ikuta and K. Kitazawa, in 
  {\it Physical Properties Of High Temperature Superconductors III}, 
  ed. D. M. Ginsberg (World Scientific, Singapore, 1992).}

\bibitem{gapexpt}{M. Itoh, S. Karimoto, K. Namekawa and M. Suzuki, 
  Phys. Rev. B {\bf 55}, 12001 (1997).}  

\bibitem{norman}{M. R. Norman {\it et al.}, Phys. Rev. B {\bf 52}, 
  615 (1995).}  

\bibitem{biplab}{B. Chattopadhyay, D. Gaitonde and A. Taraphder, 
  Europhys. Lett. {\bf 34}, 705 (1996); B. Chattopadhyay, Phys. Lett. 
  {\bf A 226}, 231 (1997); B. Chattopadhyay, J. Lahiri and A. N. Das, 
  Mod. Phys. Lett. B {\bf 11}, 1285 (1997).}  

\bibitem{evhs}{D. M. King {\it et al.}, Phys. Rev. Lett. {\bf 73}, 
  3298 (1994); Z. X. Shen {\it et al.}, Science {\bf 267}, 343 
  (1995).}

\bibitem{precur}{V. J. Emery and S. A. Kivelson, Nature (London) 
  {\bf 374}, 434 (1995); J. R. Engelbrecht {\it et al.}, 
  Phys. Rev. B {\bf 57}, 13406 (1994).} 

\bibitem{william2}{G. V. M. Williams {\it et al.}, Phys. Rev. Lett. 
  {\bf 80}, 377 (1998).}  

\bibitem{suzgrp}{M. Suzuki, S. Karimoto and K. Namekawa, J. Phys. Soc. 
  Japan {\bf 67}, 732 (1998).}  

\bibitem{pwa33}{P. W. Anderson {\it et al.}, {\it The Theory 
  of Superconductivity in the High-$T_c$ Cuprate Superconductors} 
  (Princeton Univ. Press, Princeton, 1997).}  

\bibitem{andzou}{P. W. Anderson and Z. Zou, Phys. Rev. Lett. {\bf 60}, 
  132, (1988).}    

\bibitem{fukuyama}{T. Tanamoto, H. Kohno and H. Fukuyama, J. Phys. Soc. 
  Japan {\bf 63}, 2739 (1994).}    

\bibitem{kujay}{N. Kumar and A. M. Jayannavar, Phys. Rev. B {\bf 45}, 
  5001 (1992).}

\bibitem{legget}{A. J. Leggett {\it et al}, Rev. Mod. Phys. {\bf 50}, 
  1 (1987).}

\bibitem{bilayer}{H. Ding {\it et al.}, Phys. Rev. Lett. {\bf 76}, 
  1533 (1996).}

\bibitem{dwpap1}{W. N. Hardy {\it et al.}, Phys. Rev. Lett. {\bf 70}, 
  3999 (1993); D. A. Brawner and H. R. Ott, Phys. Rev. B {\bf 50}, 
  6530 (1994); J. R. Kirtley {\it et al.}, Nature (London) {\bf 373}, 
  225 (1995).} 

\bibitem{dwpap2}{Z. X. Shen {\it et al.}, Phys. Rev. Lett. {\bf 70}, 
  1553 (1993); H. Ding {\it et al.}, Phys. Rev. B {\bf 54}, 9678 (1996).}    

\bibitem{noteeg}{In Eq.(7), $E_g$ may be accompanied by a prefactor.  
  Hence, the value of $E_g$, as used in connection with Eq.(7), may 
  differ from the actual value by a constant factor.}

\bibitem{renner}{Ch. Renner {\it et al.}, Phys. Rev. Lett. 
  {\bf 80}, 149 (1998).}  

\bibitem{unpub}{B. Chattopadhyay and A. N. Das (unpublished).} 

\bibitem{andas}{S. Sil and A. N. Das, J. Phys.: Condens. Matter {\bf 9}, 
  3889, (1997).}   

\end{thebibliography}
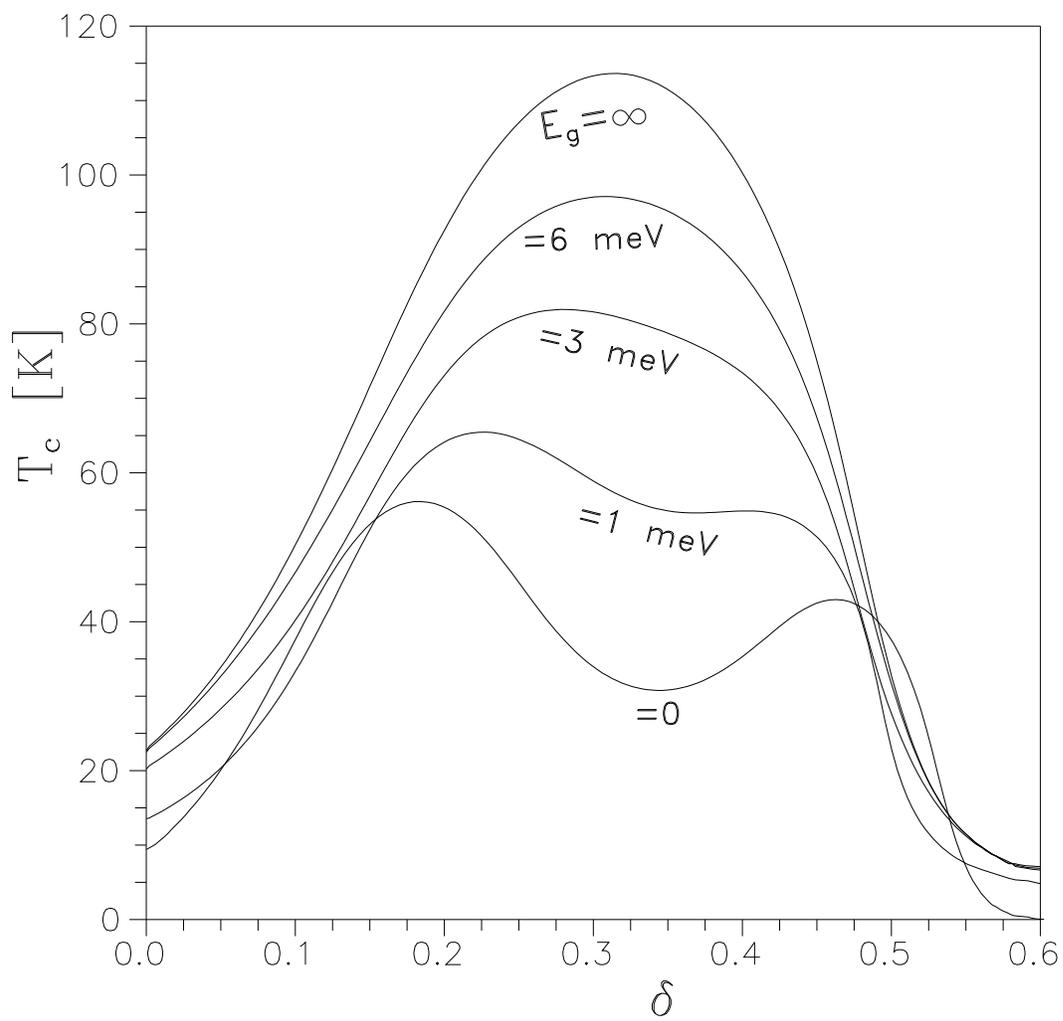
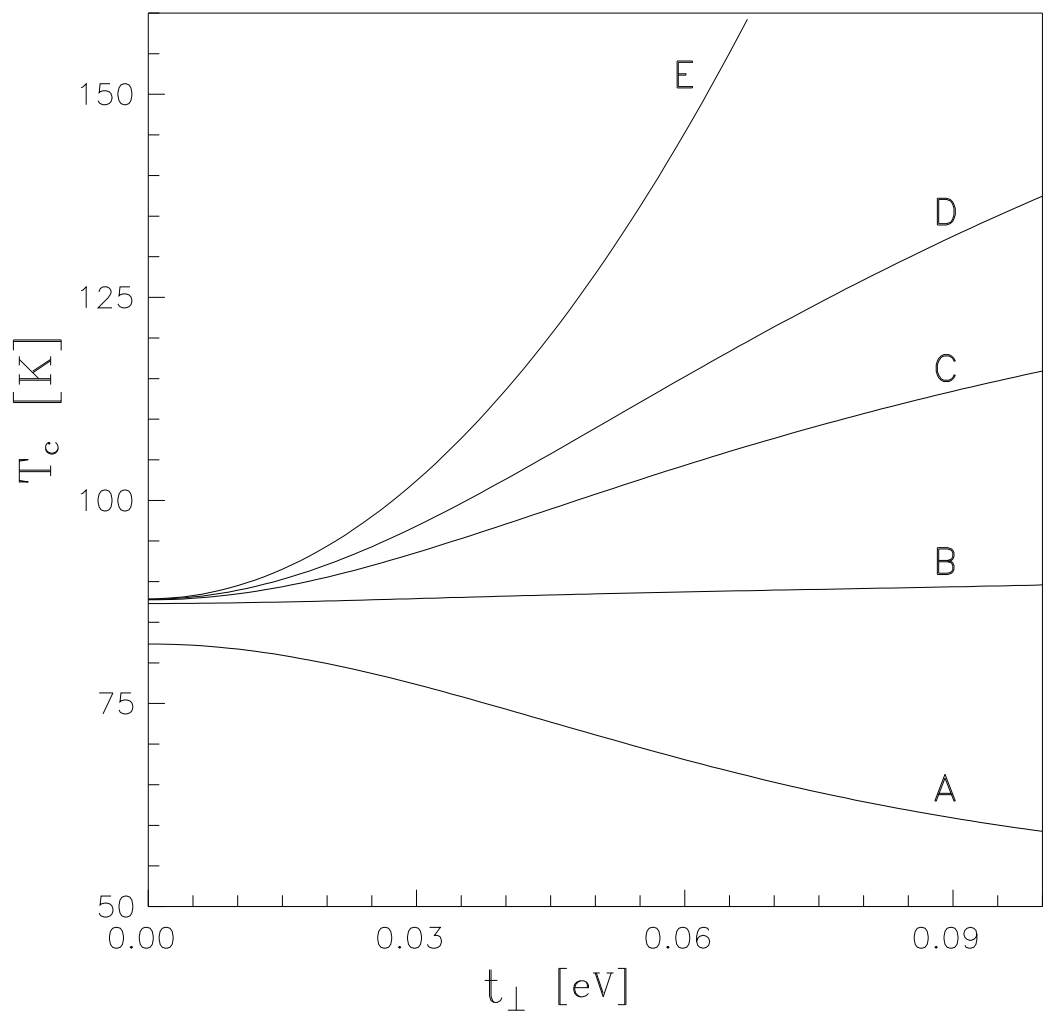
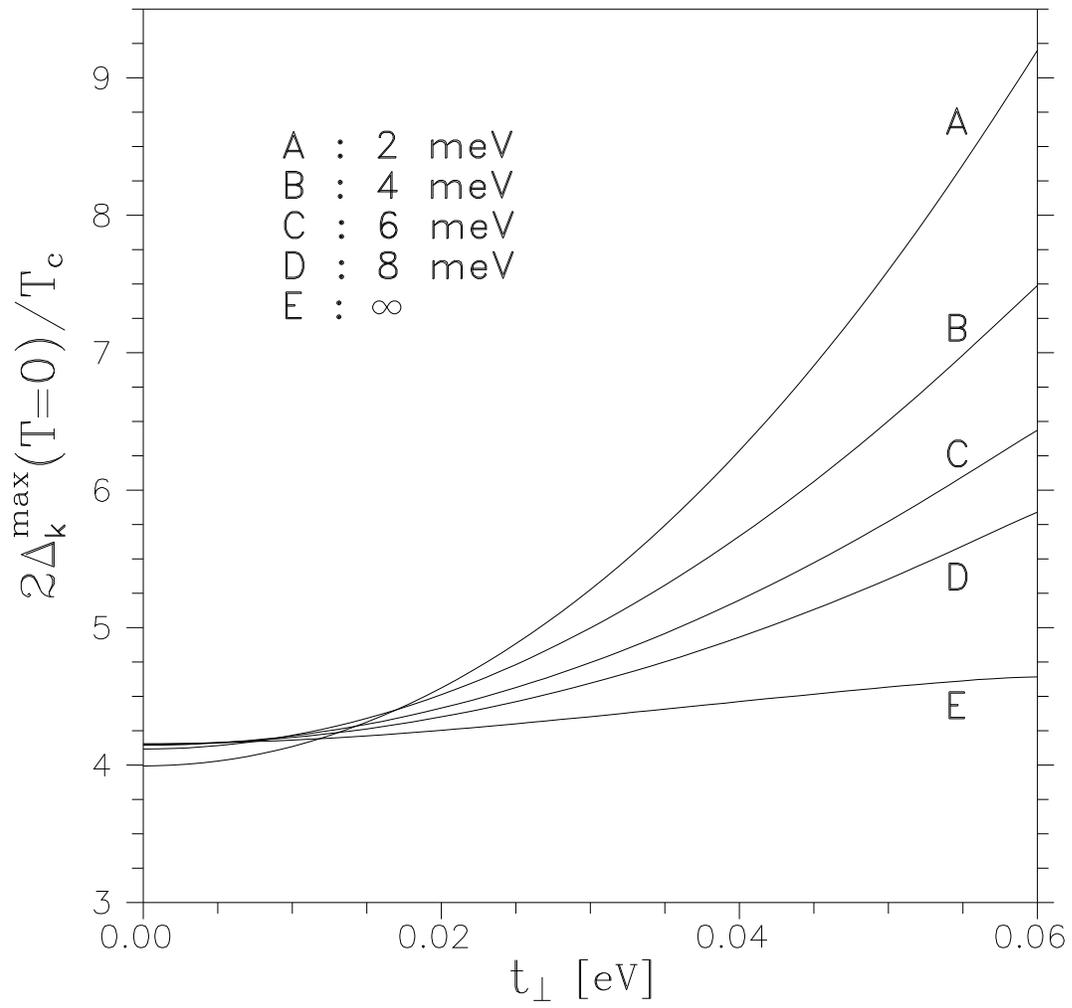
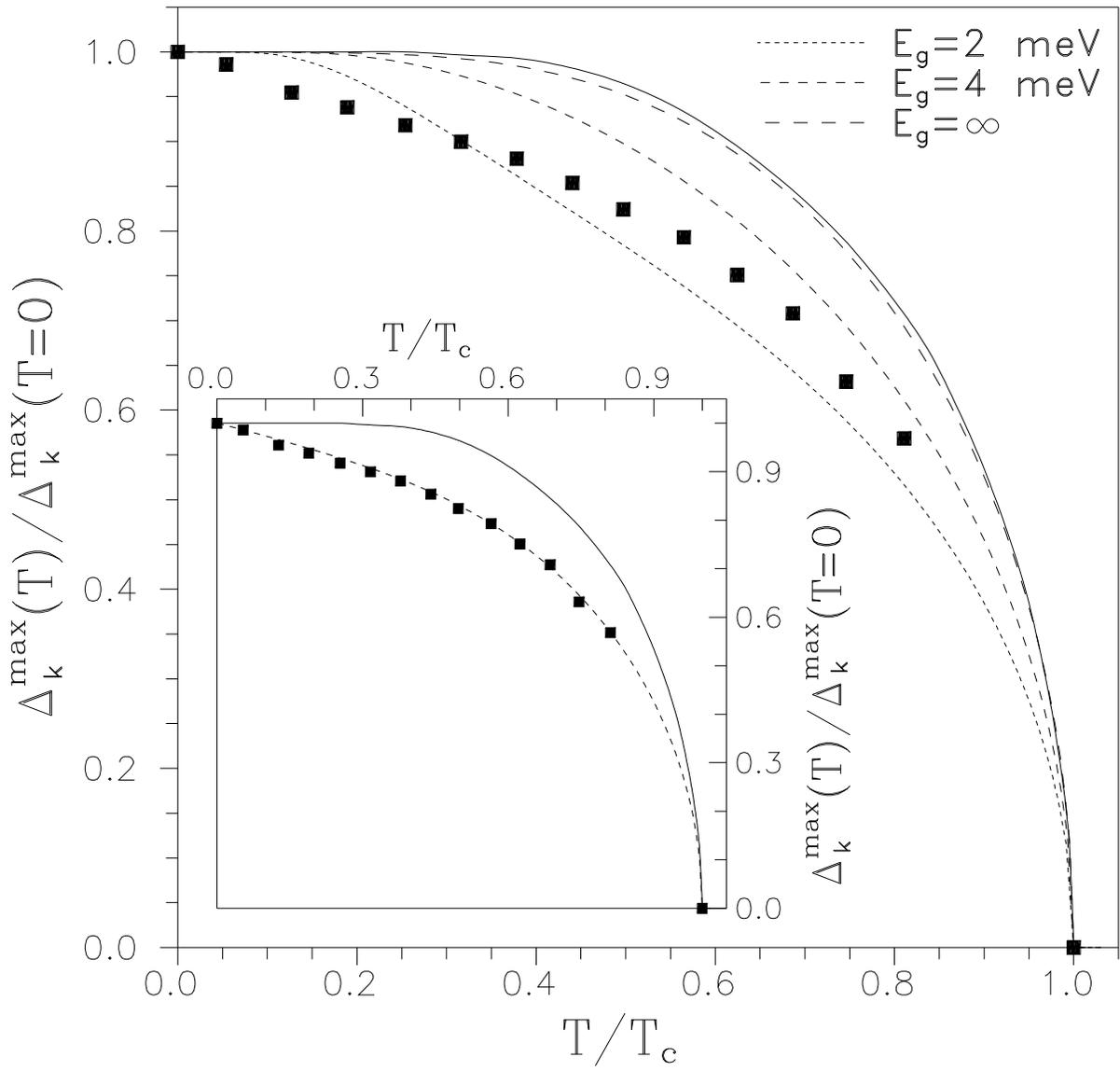
\end{document}